\documentclass{revtex4}  
\usepackage[latin1]{inputenc}
\usepackage{amsmath}
\usepackage{amsfonts}
\usepackage{amssymb}
\usepackage{graphicx,color}

\begin{document}
\title{Increase the frame rate of a camera via temporal ghost imaging}

\author{Wenjie Jiang$^{1}$, Xianye Li$^{1}$, Shan Jiang$^{1}$, Yupeng Wang$^{1}$, Zexin Zhang$^{1}$,  Guanbai He$^{1,2}$ and Baoqing Sun$^{1*}$ }
\affiliation{$^1$ School of Information Science and Engineering, Shandong University, Jinan, 250100, China\\
$^2$ State Key Laboratory of Crystal Materials, Institute of Crystal Materials, Shandong University, Jinan, 250100, China}
\date{\today}
\email{baoqing.sun@sdu.edu.cn}
\begin{abstract}
Computational temporal ghost imaging (CTGI) allows the reconstruction of a fast signal from a two dimensional detection with no temporal resolution. High speed spatial modulation is implemented to encode temporal detail of the signal into the two dimensional detection. By calculating the correlation between the modulation and the rendered image, the temporal information can be retrieved. CTGI indicates a way to detect high speed non-reproducible signal from a slow detector. Based on CTGI, we propose an innovative scheme that can increase the frame rate of a camera by resolving the temporal detail of every camera image. 
To achieve this, CTGI is conducted parallelly to different areas of the scene. High speed spatial multiplexed modulation is performed, constraining the continuous scene into a series of short-time-scale frames. All the modulated frames are accumulated into one image that is eventually used in the correlation retrieval process. By performing CTGI reconstruction on each area independently, the temporal detail of the whole scene can be obtained. This method can have a strong application in ultrafast imaging. 
\end{abstract}

\maketitle
\section{Introduction}
Modern digital cameras use focal plan array (FPA) to record two dimensional images. In visible spectrum, charge coupled device (CCD) and complementary metal oxide semiconductor (CMOS) are two typical kinds of FPAs which are most widely used. To take an image, an FPA device has to accomplish signal acquisition and data transfer consequently, which together define the frame rate of the system. Due to the limitation of detection sensitivity and data transfer speed, frame rate of normal CCD or CMOS is generally below KHz level. This frame rate is far from enough to detect high speed scene.  

To conduct ultrafast observation, there is high demand to increase the frame rate of FPA cameras. Besides the achievement in camera hardware update, computational method has become a powerful approach to break through the frame rate constraint of FPA system. Bub \textit{et al.} \cite{r1} employed a digital metal micro-mirror device (DMD) functioning as a high-speed shutter to resolve relayed images of the sample that is beyond the temporal resolution of the camera. Wilburn \textit{et al.} \cite{r2} and Agarwal \textit{et al.} \cite{r3} utilized a series of high speed triggers to control a camera array precisely and sequentially so as to achieve temporal super resolution. Pournaghi \textit{et al.} proposed a methodology of acquiring high frame rate video using multiple cameras of random coded exposure \cite{r4}. In addition, the prototype of compressive video camera has also been proposed, which takes advantage of image sparsity to conduct reconstruction from under-sampled data \cite{r6,r7}. This compressive method could increase both spatial and temporal resolution in less than 10 times.

In this work, we propose a novel scheme of high speed imaging based on temporal ghost imaging (TGI) that can capture high frame video with a slow FPA camera. Ghost imaging (GI) is an unique imaging technique that produces an image by correlating two beams of light, only one of which interacts with the object. Originally, GI was conducted in the spatial domain. It can retrieve two dimensional spatial information from non-pixelated bucket detection by utilizing spatial correlation of entangled photons \cite{r8}, classical light \cite{r9,r10}, or even computational scheme \cite{r11,r12,r13}. Spatial GI receives intensive research interests for its minimum requirement of the pixel number of the detector \cite{r14}. It was only recently that GI has been extended to the temporal domain, by taking into account space-time duality in optics \cite{r15}. Temporal ghost imaging (TGI) has been investigated theoretically, numerically and experimentally with both classical light and bi-photon state \cite{r16,r17,r18}. Same as spatial GI, TGI uses two beam correlation to retrieve an object, which enables the detection of a fast signal from a slow detector. Later on, a computational version of TGI (CTGI) is proposed based on spatial multiplexed modulation \cite{r19}. By using an FPA camera, CTGI not only simplifies TGI system into a one arm scheme, but also realizes the TGI detection in one measurement. It enables reconstruction of a single non-reproducible, periodic or non-periodic temporal signal \cite{r19,r20,r21}. The high frame rate imaging scheme proposed here utilizes CTGI to resolve temporal details of a scene captured by a camera in one exposure. To do this, we divide a camera into many identical areas. Each area composes many pixels so that it can carry out a CTGI reconstruction to resolve the temporal details of the scene projected in the area. 
By conducting CTGI in all these areas parallelly, the scheme can record videos in a speed that is much faster than the frame rate of the camera.

\section{Methods}
\subsection{Computational temporal ghost imaging}
In TGI, a test light beam with temporal fluctuation is generated and split into two arms. In the object arm, test light interacts with a temporal object and get modulated, the total power is then collected by a slow detector. Meanwhile, in the reference arm, the waveform of the test light is recorded by a fast detector. After many iteration of measurement with changing test light beams, accumulation of the two-arm correlation reveals the temporal object. Corresponding to computational spatial GI, if the test light beams are known, then TGI can be conducted in a one-arm scheme, which should be called computational TGI (CTGI). In \cite{r19}, CTGI is proposed by implementing both spatial multiplexed modulation and FPA detection. A spatial modulator displays a series of $ K$ independent random binary patterns in the size of $l\times l$  pixels. Let $ X_{k}\left ( i,j \right ) \left ( 1\leqslant k\leqslant K,1\leqslant i,j\leqslant l \right )$ be the matrix representing the modulation, where $\left ( i,j \right )$ and $k$ indicate the spacial and temporal coordinate, respectively. These successively displayed binary patterns interact with a temporal object $I(t)$ and get recorded on a multi-pixel camera in one exposure time. The process can be expressed as 
\begin{equation}
\begin{bmatrix}
S\left ( 1,1 \right )\\ 
S\left ( 1,2 \right )\\ 
\vdots\\  
S\left ( l,l \right )
\end{bmatrix}
=\begin{bmatrix}
X_{1}\left ( 1,1 \right ) &X_{2}\left ( 1,1 \right ) &\cdots\ &X_{K}\left ( 1,1 \right )\\ 
X_{1}\left ( 1,2 \right ) &X_{2}\left ( 1,2 \right ) &\cdots\ &X_{K}\left ( 1,2 \right )\\ 
\vdots \ &\vdots \  & \ddots\ & \vdots \\\ 
X_{1}\left ( l,l \right ) &X_{2}\left ( l,l \right ) &\cdots\ &X_{K}\left ( l,l \right )\\ 
\end{bmatrix}
\begin{bmatrix}
I_{1}\\ 
I_{2}\\ 
\vdots \\\ 
I_{K}
\end{bmatrix}
\label{equation1}
\end{equation}
Here $S\left ( i,j \right )$ is a two dimensional matrix representing the image recorded by the camera, i.e., the time integration of the displayed patterns modulated by the temporal object. $I_{k}$ is the transmittance of the temporal object $I\left (t\right )$ at the time when the $k$-th pattern is displayed. The temporal object can be reconstructed by calculating the intensity correlation between the time integrated image $S$ and the $K$ random patterns $X$ according to the following equation
\begin{equation}
I_{k}=\frac{\sum_{j=1}^{l}\sum_{i=1}^{l}\ (S_{ij}-\left \langle S \right \rangle)\ (X_{k}(i,j)-\left \langle X_{k}(i,j) \right \rangle)}{\sum_{j=1}^{l}\sum_{i=1}^{l}\ (X_{k}(i,j)-\left \langle X_{k}(i,j) \right \rangle)^{2}}
\label{equation2}
\end{equation}
Here $\left \langle \cdot  \right \rangle$ stands for an ensemble average for $ K$ measurements. It should be noted that the object $I_{k}$ has no spatial resolution. The final reconstructed temporal resolution is equal to the modulation steps $ K$.

\begin{figure}[h!]
\centering
\includegraphics[width=12cm]{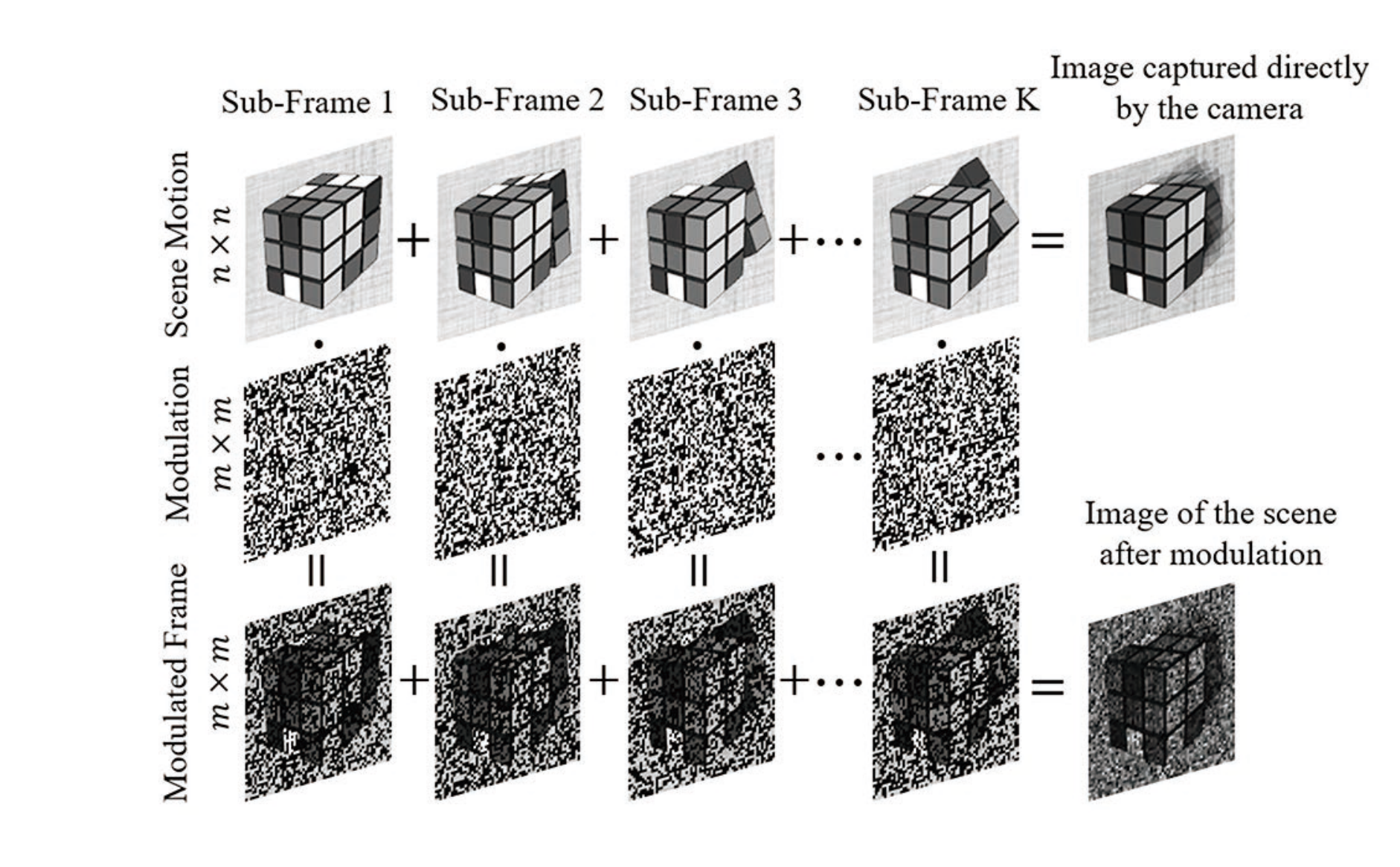}
\caption {Schematic diagram for the imaging process of the proposed scheme. The first row indicates the sub-frames that happens within one exposure time of a camera. If captured by the camera directly, movement in the scene within this process is recorded as blur without any motion detail. To resolve these sub-frames, a high speed modulation is implemented onto the dynamic scene before detection. All modulated sub-frames accumulate together as one camera image. Sub-frames are resolved by exploiting the correlation between the modulation and the rendered final image.}
\label{figure1}
\end{figure}
\subsection{High frame rate imaging scheme based on CTGI}
The fastest motion that a normal camera can record is determined by its exposure time. Motion detail within an exposure time can not be resolved and even causes smearing or motion blur. Inspired by CTGI, we propose a method to increase the frame rate of an FPA camera. The principle of our scheme can be interpreted from Fig.\ref{figure1}. To resolve the temporal information, we implement a series of $ K$ spatial modulation during one exposure. The high speed modulation split the dynamic scene into K frames. What the camera records is the accumulation of all these modulated frames. In this process, both the modulation and the camera are spatially divided into $n\times n$ "super-pixels" according to the resolution of the scene. Every super-pixel independently conducts an $l \times l$ pixels operation. In other words, for every pixel in the scene,  an area of $l \times l$ pixels both in the modulation and recording plane are allocated to perform temporal correlation measurement. By conducting CTGI reconstruction on every pixel, information of the scene in the time domain can be resolved.

In the process of spatial multiplexed operation, CTGI works to "transfer" spatial pixels into temporal resolution. The increase of imaging frame rate is therefore at the cost of spatial resolution reduction. Assuming that both the modulation and the camera are in $m\times m$ pixels, we have $ m=ln$. That is, a camera can only conduct CTGI over a scene that is $1/l^2$ of its spatial resolution. Generally speaking, given a certain reconstruction quality, more modulation pixels are required for higher temporal resolution. If the temporal resolution has been increased by $ K$ times, we define a ratio as
 \begin{equation}
T = K/ l^2
\label{equation3}
\end{equation}
$ T$ can be regarded as the transfer efficiency from a spatial pixel to its temporal counterpart. This transfer efficiency is related to the sampling efficiency according to the modulation basis $ X_{k}\left ( i,j \right )$. In \cite{r19}, random binary modulation is used. Since random binary patterns contain inherent crosstalk that will introduce inherent correlation noise and reduce reconstruction quality, to guarantee a reasonable reconstruction, the modulation resolution has to be set much larger than the temporal resolution ($T\ll1$) even in a noise-free situation \cite{r22,r23}. In our scheme, we want to minimize the spatial cost in order to realize the best imaging resolution in both time and space domain. To achieve this, we use Walsh Hadamard matrix as the modulation basis, as the orthogonality of Hadamard matrix will maximize the correlation efficiency. Taking Walsh Hadamard matrix into Eq.\ref{equation1}, one can resolve a $ K$ dimensional temporal signal in a set of complete sampling. That is, $T = 1$.

To further increase the transfer efficiency, we use compressive sensing to reduce the spatial cost of CTGI. Compressive sensing is already well studied in spatial GI or single pixel camera. It is used to reduce the measurement steps required for a reasonable single pixel image. In our scheme, we can utilize compressive sensing to reduce the spatial cost for temporal reconstruction. This is possible based on the fact that temporal signals are sparse or can be sparse in certain transform domain. To conduct compressive sensing reconstruction, the correlation measurement can be rewritten as 
\begin{equation}
y=\Phi I
\label{equation4}
\end{equation}
Where $ I$ is still a column vector with $ K$ elements that denotes the temporal object, $\Phi$ is the measurement matrix in the size of $l^2 \times K $. It means that $K$ dimensional signal $I$ is projected into an $l^2$ dimensional detection value $y$ through a certain basis $\Phi$. For compressive sensing, the size of super-pixel is less than the resolved temporal resolution, $l^{2}< K$, i.e., $T> 1$. To solve the underdetermined problem from Eq.\ref{equation4}, we adopt a Total Variation (TV) constraint term based on the fact that the gradient of most natural images is sparse. The objective function of this procedure can be written as: 
\begin{equation}
{\rm min}\ \frac{1}{2}\left \| y-\Phi x \right \|_{2}^{2}+\lambda {\rm TV}\left ( x \right ) 
\end{equation}
where the $ \left \| \cdot  \right \|_{2}$ is $l_{2}$ norm and $\lambda $ is a constant. ${\rm TV}\left ( x \right )$ is the sum of the magnitudes of discrete gradient at horizontal $D_{h}\left ( x \right )$ and vertically $D_{v}\left ( x \right )$ direction, which can be represented as ${\rm TV}\left ( x \right )=\sum \sqrt{D_{h}\left ( x \right )^{2}+D_{v}\left ( x \right )^{2}}$. There are many types of algorithms designed to solve this problem. Here, we adopt an algorithm called L1-magic \cite{r24}.

\begin{figure}[htbp]
\centering
\includegraphics[width=11cm]{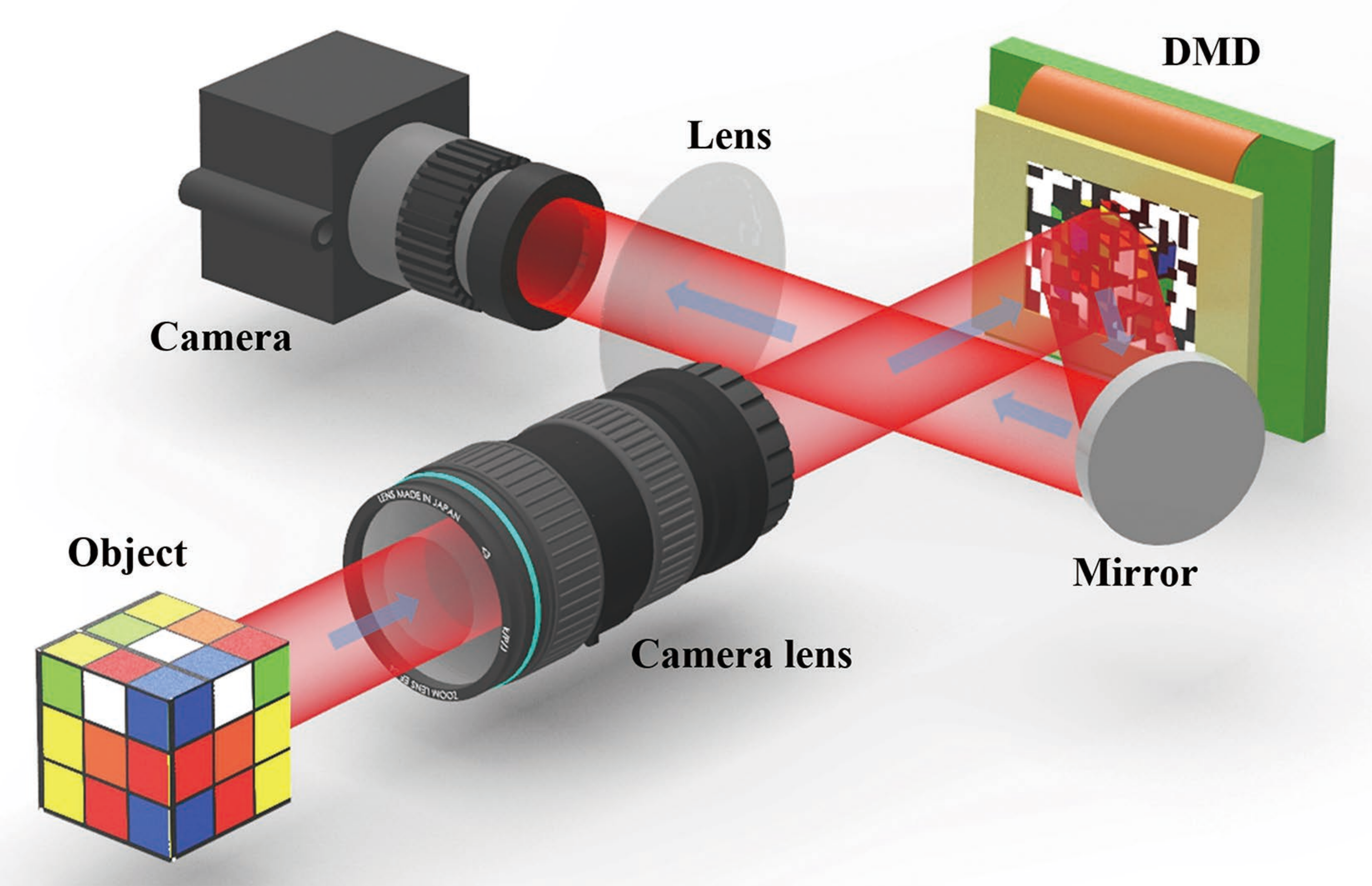}
\caption{Experimental setup of our proposed high frame rate imaging scheme. The camera lens images the object onto the DMD under ambient light. Image after modulation is then imaged and recorded by a slow camera. Both the DMD and the camera are in the same pixel resolution.}
\label{figure2}
\end{figure}

\section{Results and discussion}
\begin{figure}[htb!]
\centering
\includegraphics[width=11cm]{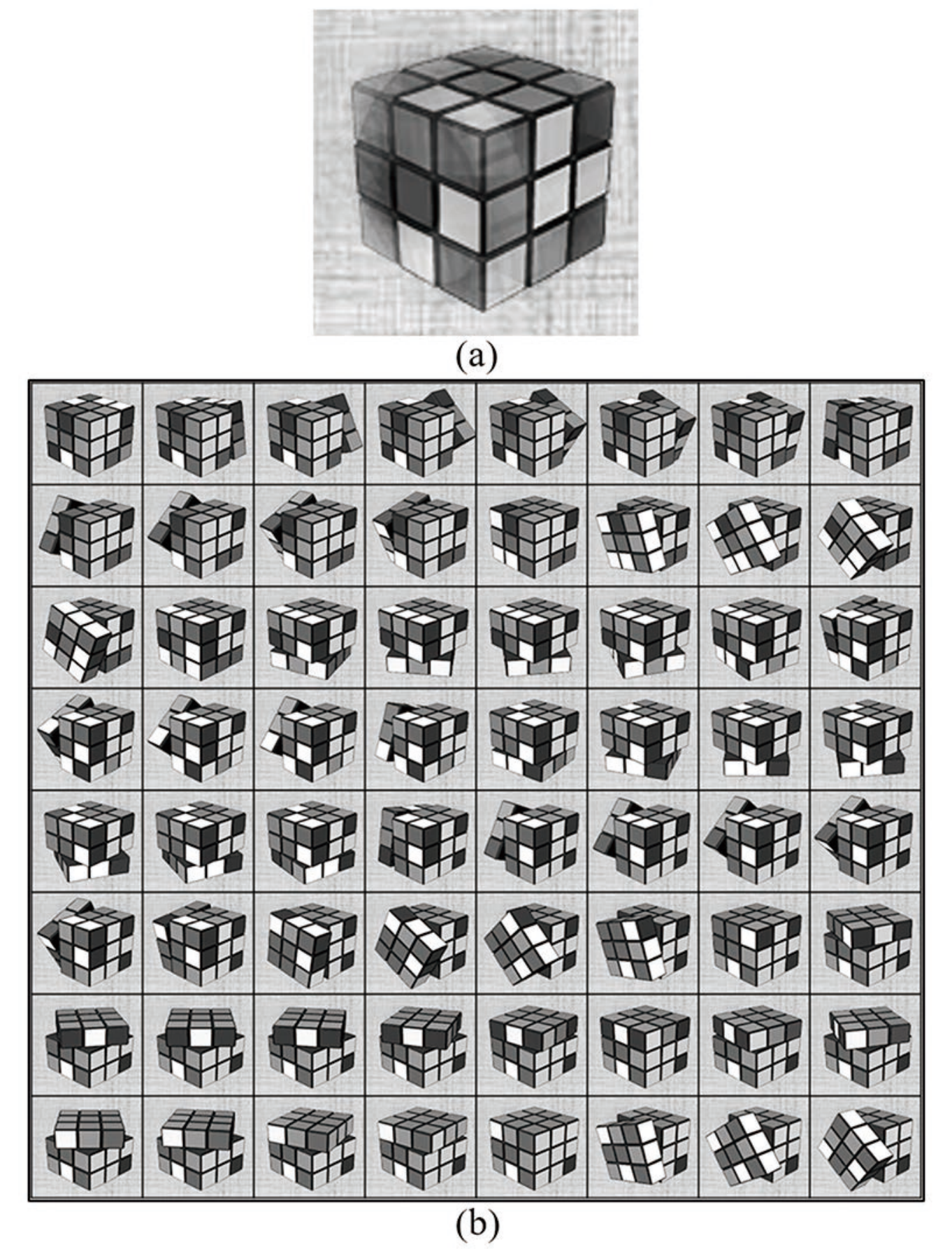}
\caption{(a) Image of the rotating rubik's cube captured by a camera directly. (b) 64 frames reconstructed from CTGI using Walsh Hadamard matrix as modulation basis.}
\label{figure3}
\end{figure}

To demonstrate the feasibility of this new high frame rate imaging scheme, proof-of-principle simulation is conducted. The system configuration is illustrated in Fig.\ref{figure2}, which includes an imaging lens, a DMD and a camera. DMD is an excellent device in the scheme for pixel multiplexed modulation on a dynamic scene, for its high spatial resolution and peak modulation frequency of over twenty kilohertz. Under ambient illumination, the scene is imaged onto the DMD by the imaging lens. To shear and modulate the incident dynamic scene, a series of $K$ binary patterns are generated and displayed on the DMD. The scene modulated by the DMD is recorded by camera at a single exposure. Our dynamic scene is about a rotating rubik's cube, composed of 64 frames of 8 bits grayscale images . The spatial resolution is $128\times128$ pixels. This 64-frame video is set to last for a period equal to the exposure time of the camera. Fig.\ref{figure3}(a) illustrates the image captured by a camera directly. The accumulation of the dynamic frames renders obvious blur and no temporal details can be seen. 

In the first simulation, to resolve the $ 64$ images from one exposure of the camera, a 64-order Walsh Hadamard matrix is selected as the modulation basis. Each of its rows (or columns) is reshaped into a two dimensional array in the size of $8\times 8$, which acts as a super-pixel of the modulator. A total array of $128\times 128$ super-pixels are employed to match the spatial resolution of the target. A total number of $ 64$ steps of modulation are conducted. The entire modulation basis can be represented as a three dimensional array in the size of $128\times 128\times 64$ super-pixels. Mathematically, the modulation process is the Hadamard product between the modulation basis and the dynamic scene. All the modulated frames are added together to form a camera image, which is in the same spatial resolution as the DMD, i.e., $1024\times 1024$ pixels. From this image together with the modulation basis, a 64-frame video can be reconstructed using Eq.\ref{equation2}. Fig.\ref{figure3}(b) shows all the frames reconstructed by our high frame rate imaging scheme, and its temporal resolution is 64 times that of Fig.\ref{figure3}(a).

\begin{figure}[htbp]
\centering\includegraphics[width=11cm]{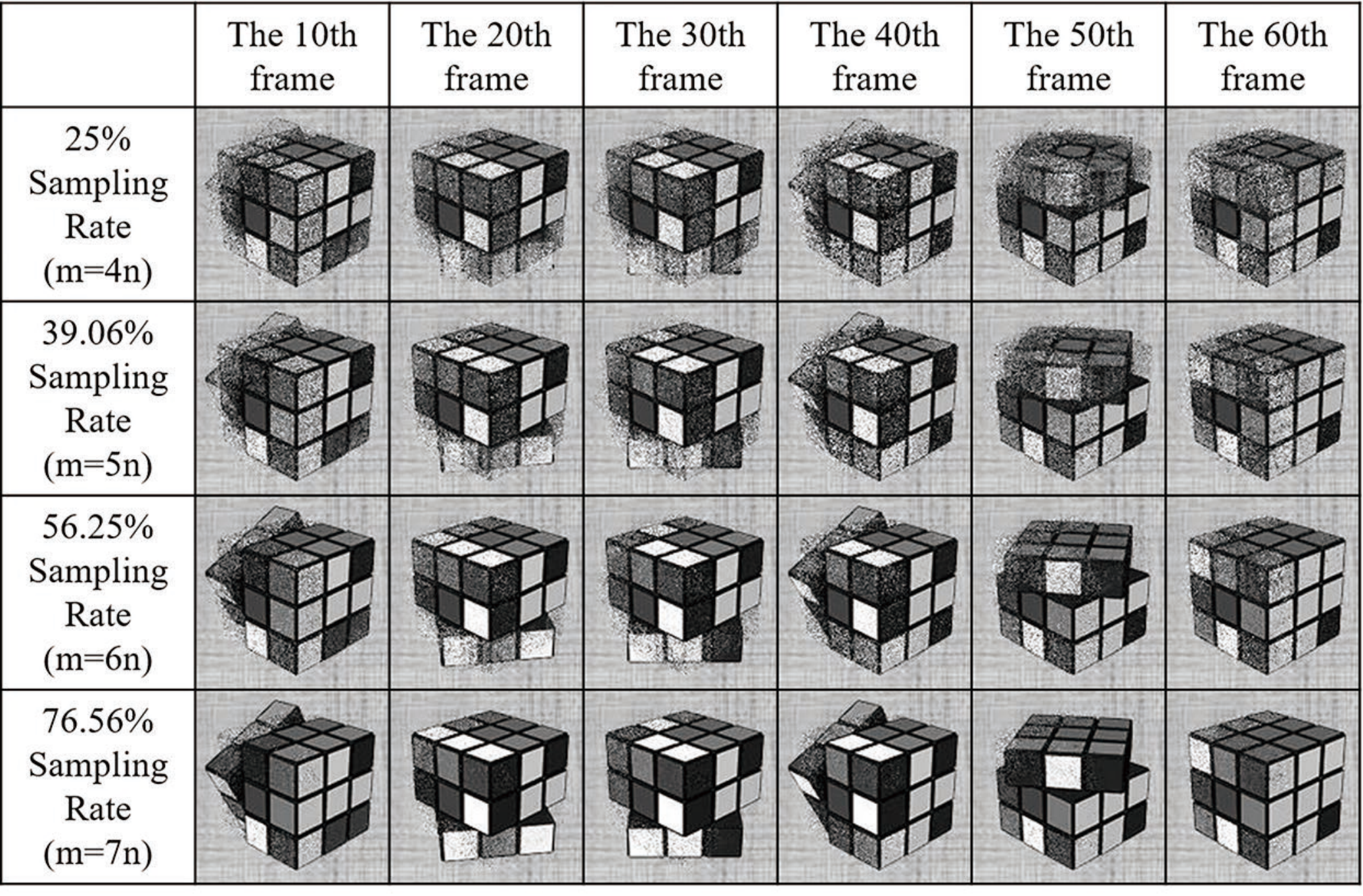}
\caption{Selected reconstruction frames using compressive sensing algorithm at different sampling rates. Reconstruction are performed under four different sampling rates, i.g., 25\%, 39.06\%, 56.25\%, 76.56\% respectively.}
\label{figure4}
\end{figure}

In our second simulation, the compressive sensing algorithm is employed. To minimize the sampling rate, instead of specially designed matrix, random binary patterns are used as the modulation basis. As compressive sensing enables reconstruction from under-sampled data, the spatial resolution of a super-pixel is less than $ 8 \times 8$ and varies according to the sampling rate. Fig.\ref{figure4} shows selected CTGI frames under four different sampling rate. Low sampling rate means less spatial cost for certain temporal resolution. On the other hand, less sampling rate will lead to worse image quality. For example, at a sampling rate of $ 25\%$, i.e., $T=4$, the required effective resolution of the camera and modulator is $512\times 512$ pixels, a quarter of that for Hadamard reconstruction. However, the reconstructed images, especially the moving part, are severely dimmed. As the sampling rate increases, the quality of reconstructed images improves significantly. When the sampling rate reaches 76\%, i.e.,$T=1.3$, a nearly perfect reconstruction can be achieved. In practical application, the sampling rate for an acceptable reconstruction varies dependent on the spatial and temporal complexity of the motion. For some simple moving targets, one could achieve a highly compressed reconstruction at low spatial cost.

\begin{figure}[h!]
\centering
\includegraphics[width=11cm]{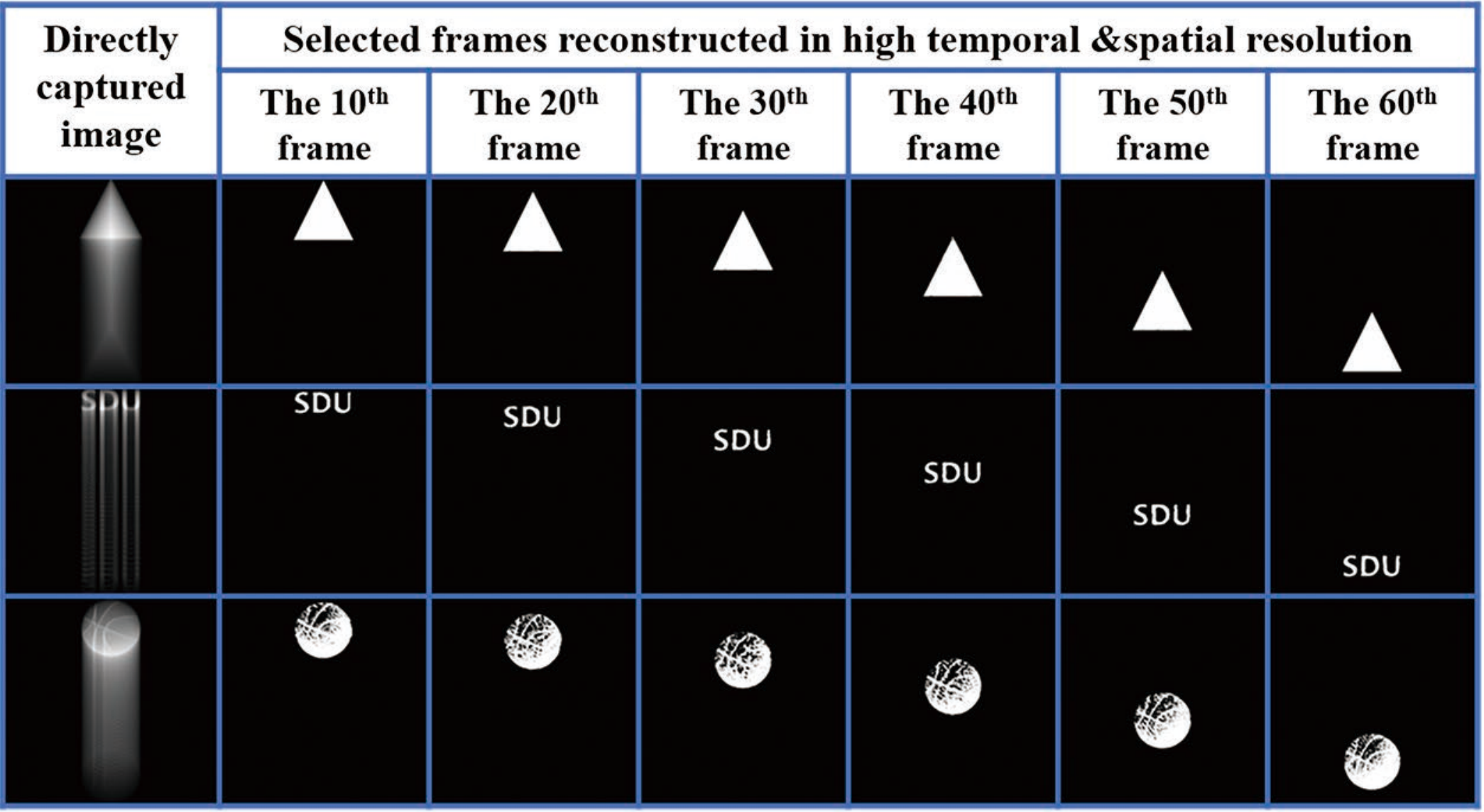}
\caption{High frame rate reconstruction by exploiting the periodic structure of the modulation for high reconstruction resolution. Three objects are used as the target, including a binary triangles, a binary pattern composed of letters ''SDU'' and a grayscale basketball. The scene is about the free fall process for a period equal to the time of one exposure of the camera. The left column are the images captured by a camera directly. Figures on the right shows selected frames obtained from CTGI. By conducting CTGI reconstruction based on any $ 8 \times8$ pixels in the modulation plane, a reconstruction resolution of  $ 1017 \times 1017$ pixels is achieved from a $ 1024 \times 1024$ pixels modulation.}
\label{figure5}
\end{figure}

Both simulations above are conducted in a way that each target pixel is projected onto an exclusive modulation area. This is to ensure that the target within a super-pixel is uniform at each modulation step, as non-uniform distribution within a super-pixel will lead to crosstalk noise between different pixels and different frames. In practice, however, due to the complexity of object as well as the limited spatial resolution of the modulation device, it is more likely to have a non-uniform distribution within a super-pixel, which then neglects the necessity for independent modulation. Standing at this point, in the Hadamard scheme, we can take advantage of the periodic structure of the spatial modulation to maximize the reconstruction resolution. More specifically, in the situation above, each spatial modulation is an array of $128\times 128$ duplication of certain $ 8\times 8$ pixels pattern. Therefore, any adjacent $ 8 \times 8$ pixels on the DMD then form a complete modulation basis from 64 modulation steps. In that case, the reconstructed frames are in a scale of $ 1017 \times 1017$ pixels. A simulation applying this approach is shown in Fig.\ref{figure5}. The dynamic target is about the free fall of three simple objects in a constant background, each being represented in a 64-frame video. The left column shows the image captured by a camera in one exposure. The resolved images at different frames are shown on the right part of the figure. All the reconstructed images in Fig.\ref{figure5} are in a spatial resolution of $1017 \times 1017$ pixels. As mentioned above, the non-uniform distribution of the target within a super-pixel leads to crosstalk noise and eventually trailing effect in the resolved frames. This trailing effect has been removed in the results simply by adding a threshold to the reconstructed images. 

\section{Conclusion}
To summarize, we have proposed a novel method to boost the frame rate of a camera based on CTGI. By implementing high speed spatial multiplexed modulation in front of a camera detection, the temporal information of a camera image can be resolved. The boost of frame rate is at the cost of spatial resolution, and dependent on the modulation efficiency. To achieve a reasonable temporal resolution while minimizing the loss of spatial resolution, two reconstruction algorithms, iterative Hadamard correlation and compressive sensing, are employed for proof-of-principle simulation demonstration. To further minimizing the spatial cost, in the Hadamard scheme, we take advantage of the periodic structure in the modulation to form a maximum number of CTGI modulation unit by combining any adjacent modulation pixels in certain size. Our high frame rate scheme is of great application potential, especially for low spatial resolution circumstances, i.g., to track the location of a fast-moving object. Using a high speed spatial modulator such as DMD, one can easily increase the frame rate of a slow camera in a magnitude of two or even higher. 
\section*{Funding}
This work was supported by National Natural Science Foundation of China (NSFC) (Grant No. 61675117).
\bibliography{reference}
\end{document}